\documentstyle[preprint,aps,12pt]{revtex}
\begin{document} 
\date{3 December 1997}
\title{Photon assisted electric field domains  
in doped semiconductor superlattices} 
\author{Ram\'on Aguado and Gloria Platero}
\address{Instituto de Ciencia de Materiales (CSIC) Cantoblanco,28049
Madrid, Spain.}
\maketitle
\begin{abstract}
We study photon-assisted
tunneling (PAT) through weakly-coupled doped semiconductor superlattices (SL's)
in the high voltage regime.
A self-consistent model which treats the Coulomb interaction in a
mean field approximation is considered.
We discuss the formation of electric field domains in the presence of THz
radiation and the appearance of new multistability regions caused by
the combined effect of the strong non linearity coming from the Coulomb 
interaction and the new PAT channels.
We show how the electric field domains can be
supported by the virtual photonic sidebands due to multiple photon
emission and absorption.
Ramon Aguado, ICMM(CSIC), Cantoblanco 28049, Madrid, Spain.\\
Fax: 34-1-3720623, 
e-mail: raguado@everest.icmm.csic.es
\end{abstract}
Recently, the transport properties of semiconductor
nanostructures in the presence of AC fields have
been the subject of interest \cite{1,2,3}. Among the experiments based on the 
response of a nanostructure to an AC field one might
consider PAT, electron pumps, turnstiles, 
and others .
On the other hand the transport properties having their origin in
Coulomb interaction have
attracted a great deal of attention in recent times. In
weakly-coupled SL's, multistability, electric field domain
formation, self-sustained current oscillations and chaos have been much
studied \cite{4,5}.
In this work we deal with PAT in weakly-coupled doped SL's 
whose transport mechanism is sequential
tunneling. In this regime the intrinsic miniband width is
typically one order of magnitude smaller than the scattering induced
broadening (for the samples we are going to analyze the miniband width
is $\sim 0.1 meV$). In this case the miniband transport can be
neglected and the transport at low bias voltages is governed by
sequential resonant tunneling from ground state to ground state 
in adjacent wells.
The PAT through a SL in the low bias regime, 
where the non linear effects of the
Coulomb interaction are small, has been explored both
experimental \cite{1} and theoretically \cite{6}.
In this regime, and for certain values of the
intensity and frequency of the AC field, the
sample can exhibit negative conductance \cite{1,6}. 
Here, we concentrate in the high bias regime where the
Coulomb
interaction has to be considered in order to discuss electric field
domains formation in the presence of THz fields. 
We study the formation of new multistability regions induced
by the AC field and we show how the electric field domains can be
supported by the virtual PAT sidebands \cite{7}.
An extension of the model of Ref. 10 is put forward in order to account
for the effect of an AC field on the I--V
curve of a doped SL.
The model assumes that the characteristic time of
intersubband relaxation is shorter than the
tunneling time and then
that only the ground state of each quantum well is
populated.
In order to model relaxation in the transport direction
due to scattering , we assume 
that the spectral functions in the wells are Lorentzians.  
Relaxation in the
planes perpendicular to the transport is taken into account by imposing current
conservation through the whole structure that will determine
the sequential current as well as the Fermi energies within each
well.
The assumption of having Fermi distribution functions in the
wells implies some scattering mechanism that
thermalizes
the electrons towards an equilibrium state \cite{8}. If there
is no mechanism for obtaining thermal equilibrium
the nonequilibrium transport must be
treated with a suitable quantum kinetic technique (for example the
Keldysh technique \cite{3}).\\
The currents are calculated by means of the Transfer Hamiltonian
method (TH) that consists in connecting by time dependent
perturbation theory the different parts of the
structure (decoupled in the remote past) by tunneling matrix
elements \cite{9}.\\
The effect of the AC 
field is included in 
the single particle energies in the different isolated regions of
the structure (leads and wells);
$\epsilon_{i}(t)=\epsilon_{i}+eV_{i}(t)$ 
before the tunneling couplings are switched on \cite{3}. 
Due to this time dependence the
single particle propagators acquire phase factors. The 
noninteracting retarded Green's function at the $i$-th region becomes:
\begin{eqnarray}
g^{r}_{i}(t,t')&\equiv& -\frac{i}{\hbar}\theta(t-t')\langle\{{\bf
c}_{k_{i}}(t),{\bf c}^{\dag}_{k_{i}}(t')\}\rangle=-\frac{i}{\hbar}
\theta(t-t')e^{[-\frac{i}{\hbar}\int_{t'}^{t} d\tau
\epsilon_{i}(\tau)]}\nonumber\\
&=&\sum_{n,m=-\infty}^{\infty}
J_{n}(\frac{eV_{i}}{\hbar \omega}) J_{m}(\frac{eV_{i}}{\hbar
\omega})
e^{-\frac{i}{\hbar}\epsilon_{i}(t-t')}e^{-i n\omega t} e^{i
m \omega t'} . 
\end{eqnarray}
In this expression $J_{n}$ is the Bessel function of first kind
and the time dependent voltage is
$V_{i}(t)=V_{i}\cos\omega t$.
Following the TH method 
we obtain the transmission
probability from the $i$-th well to the $i+1$-th well 
, from the emitter to the first well and from the $N$-th well ($N$ is the
number of wells) to the
collector. From the transmission probability we
evaluate the time averaged sequential current .
The interwell current is \cite{6}:
\begin{eqnarray}
J_{i,i+1}&=&\frac{2e\hbar
k_{B}T}{\pi^{2}m^{*}}\sum_{j=1}^{n_{max}}\sum_{m=-\infty}^{\infty}
J_{m}^{2}(\frac{eFd}{\hbar \omega})
\int\frac{\gamma}{[(\epsilon-\epsilon_{C1}^{i})^{2}+\gamma^{2}]}
\frac{\gamma}{[(\epsilon-\epsilon_{Cj}^{i+1}+m\hbar\omega)^{2}+\gamma^{2}]
}\nonumber\\
&\times& T_{i+1}(\epsilon,\epsilon+m\hbar\omega)
ln [\frac{1+e^{\frac{(\epsilon_{\omega_{i}}-\epsilon)}{k_{B}T}}}
{1+e^{\frac{(\epsilon_{\omega_{i+1}}-\epsilon-m\hbar\omega)}{k_{B}T}}}
] d\epsilon ,
\end{eqnarray}
where $\epsilon_{Cj}^{i}$ 
is the $j$-th resonant state of the $i$-th well ($n_{max}$ is the
number of subbands participating in the transport)
, $T_{i}(\epsilon,\epsilon+m\hbar\omega)$
is the inelastic transmission through the $i$-th barrier. 
In the argument of the Bessel functions , 
$Fd=V_{i+1}-V_{i}$ is the potential drop
between the $i$-th well and the $i+1$-th well due to the time
dependent field; $F$ the intensity of the time dependent 
external field and $d$ the
period of the SL.
The current
from the emitter to the first well, $J_{0,1}$,
and the one from the $N$-th well to the collector, $J_{N,N+1}$, 
are also derived in our model.
The electrons in each well are in local
equilibrium with Fermi energies $ \epsilon_{\omega_{i}} $ which
define the charge densities $n_i$.
For a given set of variables $\{\epsilon_{\omega_{i}}\}$ 
and in the stationary
regime the currents have to fulfill the set of equations \cite{10}:
\begin{equation}
\ J_{i-1,i}-J_{i,i+1}=0
\hspace{2cm} i=1,\ldots,N.\label{rate}
\end{equation} 
Here $J_{i,i+1} = J_{i,i+1}(\epsilon_{\omega_{i}},
\epsilon_{\omega_{i+1}},\Phi)$,
$J_{0,1} = J_{0,1}(\epsilon_{\omega_{1}},\Phi)$, and $J_{N,N+1} =
J_{N,N+1}(\epsilon_{\omega_{N}},\Phi)$. $\Phi$ denotes the set of
non linear voltage drops (potential drops at the
accumulation and depletion layers, barriers and wells)
through the structure caused by the
accumulation of charge densities $n_i$ in the wells.
These potential drops as well as all the quantities in the problem
are calculated
self-consistently for each applied voltage including the Coulomb
interaction in a mean field approximation \cite{10}. Our numerical
procedure allows us to obtain both the stable and unstable solutions as
well as all the multistability regions in the I--V curve.\\
In Fig. 1 we plot the I--V curve of a SL consisting in 10 wells with 
$90 \AA {\rm GaAs}$ wells and $40 \AA  {\rm Ga_{0.5}Al_{0.5}As}$ barriers. 
The doping at the leads is $N_{D}=2\times 10^{18} {\rm cm^{-3}}$ 
and in the wells it
is $N_{D}^{w}=1.5\times 10^{11} {\rm cm^{-2}}$, the half-width of
the resonant states is $\gamma=2 {\rm meV}$ and T=0.\\ 
After the initial peak which is determined by C1--C1 sequential tunneling
($C_{j}$ are the resonant states of the wells ordered starting from the ground
state)
through the whole SL the current evolves along a series
of branches (solid lines). 
This behaviour is explained by the formation of a
charge accumulation layer in one of the wells (domain wall)\cite{4,5} that splits the
SL in two regions with low and high electric field respectively 
(see inset in Fig. 2).
Increasing the voltage, this charge cannot move continuously 
through the SL. This motion can only
occur for voltages allowing resonant interwell tunneling, in this situation
the domain wall moves from the $i$-th well to the $i-1$-th well. In the
I--V characteristics this process leads to a series of sharp
discontinuities (the stable branches are connected by unstable ones,
dotted lines in the figure).
In the static
case, transport in the high electric field domain 
is only possible by C1--C2 resonances, whereas in
the case with AC the resonances occur between the virtual PAT emission
sidebands corresponding to the C2 subband in the
$i+1$-th well and the PAT absorption sidebands 
corresponding to the C1 subband  
in the $i$-th well .
These processes are shown in Fig. 2 where the I--V curve in the
presence of THz radiation ($F=0.47\times 10^{6} {\rm V/m},\omega=3 {\rm
THz}$) 
is plotted. 
In the inset we have
depicted the calculated potential profile of
 the SL for a voltage $V_{0}=0.86 {\rm V}$. The
inset shows that the virtual PAT channels can support electric field
domains (the dashed lines represent the PAT channels), the blow up
shows that the high field domain is supported by C1--C2 tunneling
involving absorption of two photons. 
Increasing the intensity of the AC field to $F=0.95\times
10^{6} {\rm V/m }$ (Fig. 3) the probability of having multiphotonic
effects increase, leading to multistability of the branches. The
inset shows a magnification of the first branch, the circles mark the stable
operating points for a fixed voltage. At $V_{1}=0.16{\rm V}$
transport in the high field domain 
occurs via tunneling between the two-photon absorption
virtual state associated with C1 and the two-photon emission
virtual state associated with C2. At $V_{2}=0.19{\rm V}$ the branch
develops a multistable solution (five solution coexist, three stable,
two unstable).  
These solutions correspond to a different number
 of photons emitted in C2: one photon in the
highest current stable solution (circle a), two photons in the lowest current 
stable solution (circle c);
the process from the highest current to the lowest one involves the
motion of the domain wall. The situation
repeats periodically as the domain wall moves, giving the sawtooth structure in
the current.\\ 
In summary, we have presented and solved a microscopic self-consistent model for the
sequential current through a weakly-coupled doped SL in the presence of THz
radiation. We have shown how the electric field domains can be supported by the
virtual PAT sidebands. For strong THz fields multiphotonic effects lead to
appearance of new multistability regions in the I--V
curve.\\
We acknowledge L. L. Bonilla and M. Moscoso for fruitful discussions and
collaboration on related topics. This work has been supported by the
CICYT (Spain) N0.MAT 94-0982-c02-02.

\begin{figure}
\caption{ I--V characteristic for a SL without AC. The continuous (dotted)
lines correspond to stable (unstable) solution branches.}
\end{figure}
\begin{figure}
\caption{ I--V characteristic for a SL with AC
($F=0.42\times10^{6} V/m,\omega=3THz$).The inset shows the
calculated potential profile at V=0.86 V.}
\end{figure}
\begin{figure}
\caption{ I--V characteristic for a SL with AC
($F=0.95\times10^{6} V/m,\omega=3THz$).The inset shows a blow up of the
first branch.}
\end{figure}
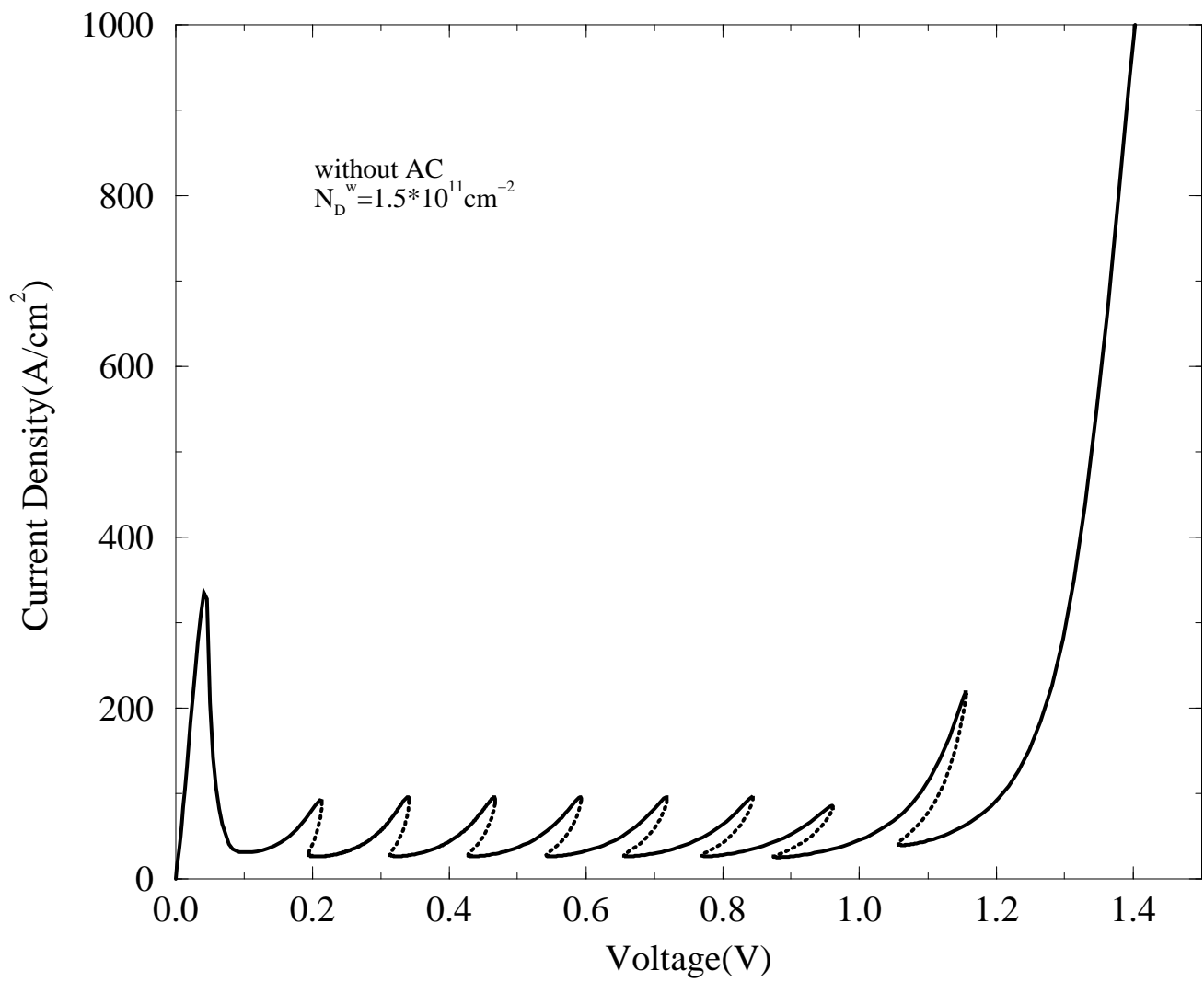
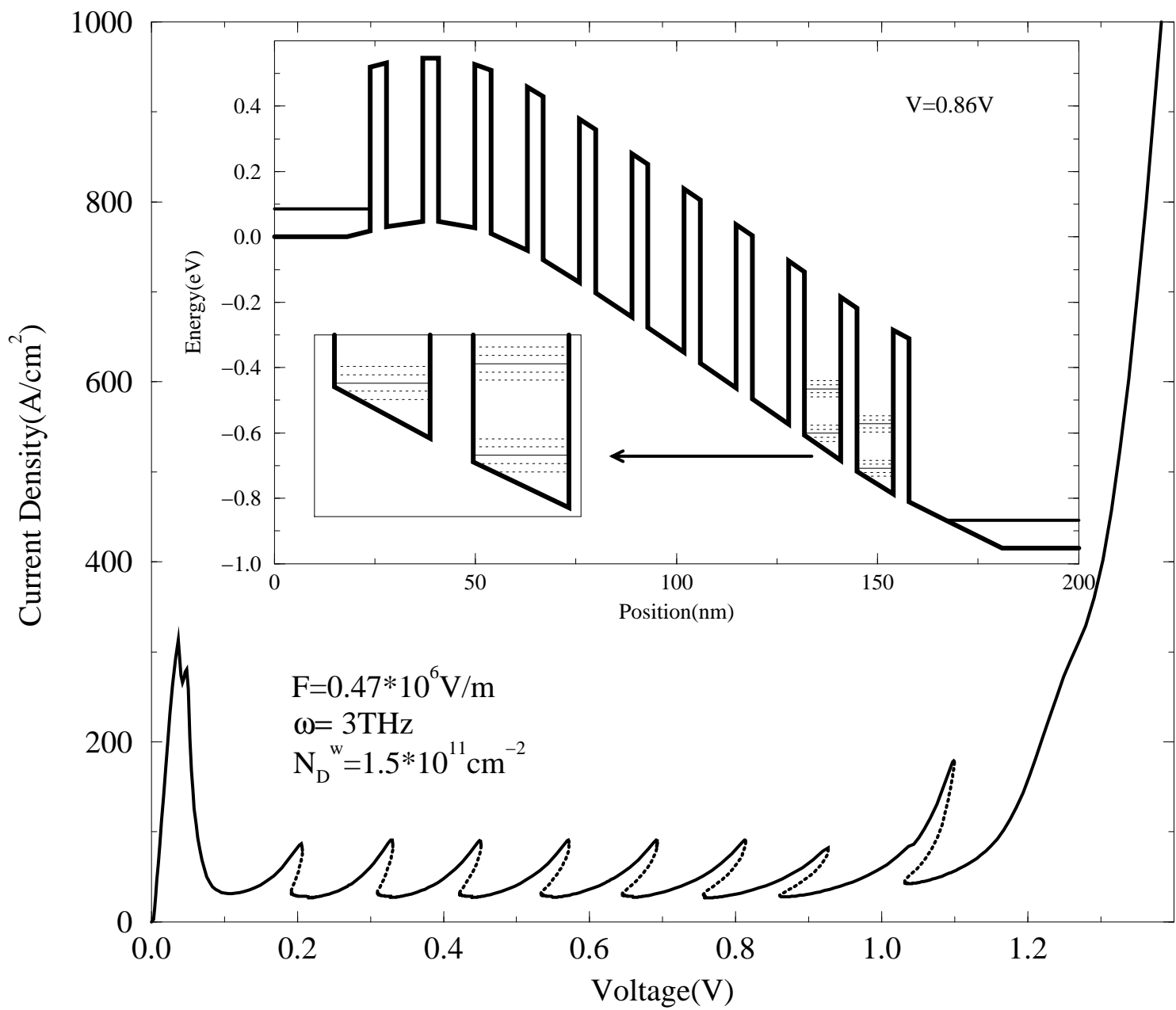
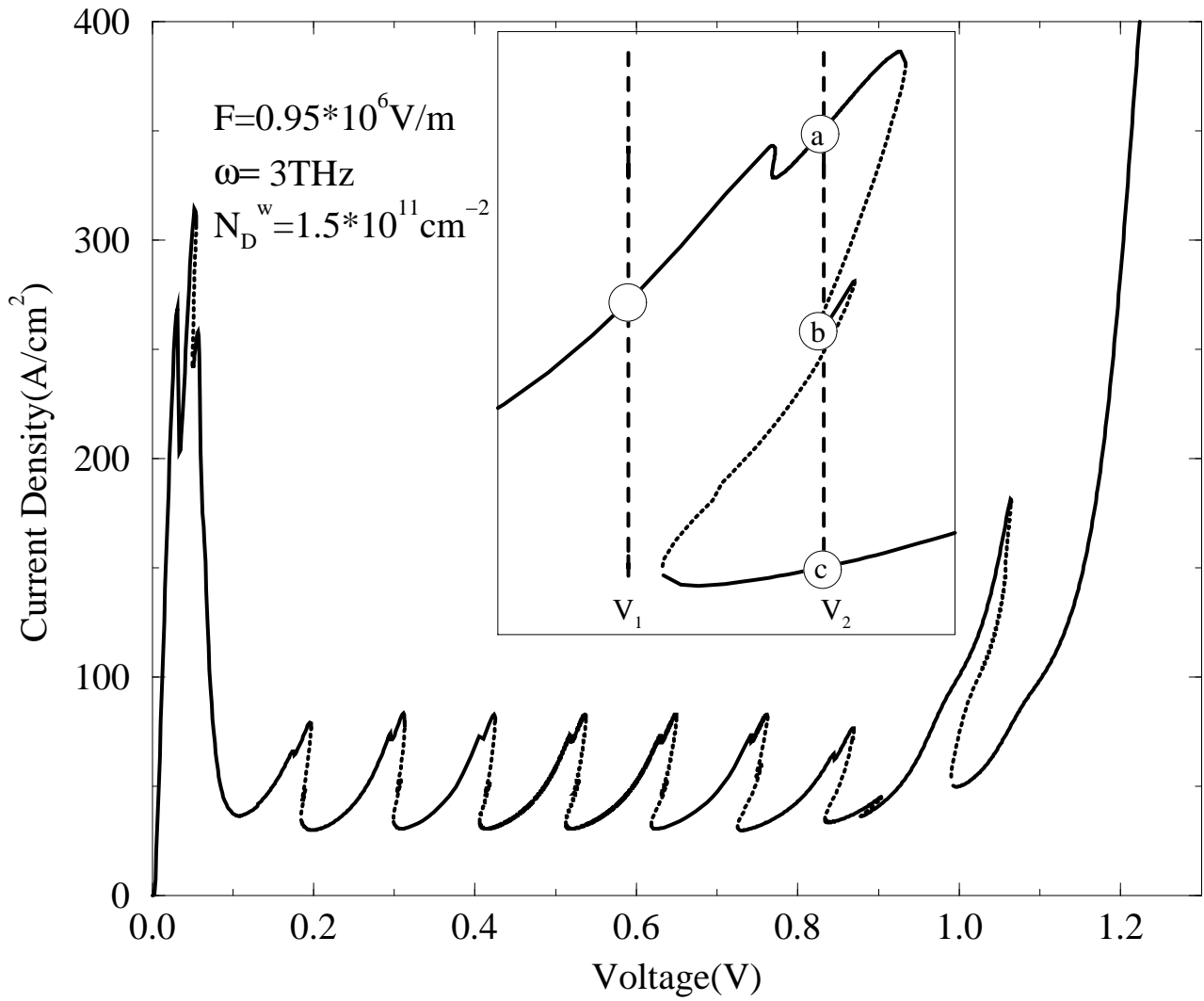
\end{document}